\documentclass[11pt]{article}

\usepackage{amsmath}
\usepackage{amsfonts}

\usepackage{hyperref}
\hypersetup{
   colorlinks   =  true,
    linkcolor    = cyan,
    citecolor    = red,
     urlcolor	=magenta,     
}

\newcommand\be{\begin{equation}}
\newcommand\ee{\end{equation}}
\newcommand\bea{\begin{eqnarray}}
\newcommand\eea{\end{eqnarray}}

\def\t{\vartheta}
\def\r{\eta}
\def\L{\Lambda}
\def\p{\beta}
\def\pp{\alpha}

\def\s{s}
\def\ro{\eta}
\newcommand\dd{{\rm d}}
\newcommand\aaa{A}
\newcommand\bbb{B}

 \def\>#1{{\mathbf #1}}                 

\newcommand\esf{{\mathcal S}} 

\begin{document}

\begin{center}
\baselineskip 24 pt {\LARGE \bf  
The $\kappa$-(A)dS noncommutative spacetime} 
\end{center}

\bigskip 
\medskip

\begin{center}

{\sc Angel Ballesteros, Ivan Gutierrez-Sagredo, Francisco J.~Herranz}

{Departamento de F\'isica, Universidad de Burgos, 
09001 Burgos, Spain}

e-mail: {\href{mailto:angelb@ubu.es}{angelb@ubu.es}, \href{mailto:igsagredo@ubu.es}{igsagredo@ubu.es}, \href{mailto:fjherranz@ubu.es}{fjherranz@ubu.es}}

\end{center}

\begin{abstract}
The  (3+1)-dimensional $\kappa$-(A)dS noncommutative spacetime is explicitly constructed by quantizing its semiclassical counterpart, which is the $\kappa$-(A)dS Poisson homogeneous space. This turns out to be the only possible generalization  of the well-known $\kappa$-Minkowski spacetime to the case of non-vanishing cosmological constant, under the condition that the time translation generator of the corresponding quantum (A)dS algebra is primitive. Moreover, the $\kappa$-(A)dS noncommutative spacetime is shown to have a quadratic subalgebra of local spatial coordinates whose first-order brackets in terms of the cosmological constant parameter define a quantum sphere, while the commutators between time and space coordinates preserve the same structure of the $\kappa$-Minkowski spacetime. When expressed in ambient coordinates, the quantum $\kappa$-(A)dS spacetime is shown to be defined as a noncommutative  pseudosphere.
\end{abstract}

\noindent
PACS:   \quad 02.20.Uw \quad  03.30.+p \quad 04.60.-m

\bigskip

\noindent
KEYWORDS: quantum groups, cosmological constant, (Anti-)de Sitter, kappa-deformation, noncommutative spacetimes, quantization


 \section{Introduction}

Noncommutative spacetimes and their associated uncertainty relations are widely expected to provide suitable frameworks for the description of minimum length or fuzziness features of the spacetime arising in different approaches to Quantum Gravity (see, for instance,~\cite{Snyder1947,Yang1947,Maggiore1993gup,Maggiore1993algebraicgup,DFR1994,DFR1995,Garay1995,Hossenfelder2013minlength} and references therein). As a consequence, several notions of noncommutative spacetimes have been proposed in an attempt to describe this ``quantum geometry", and a remarkable common feature of all these approaches is a shift from geometry to algebra~\cite{Connes1994NCGbook}. In particular, when considering noncommutative spacetimes arising from quantum groups, the emphasis is put in the introduction of some (deformed/quantum) symmetry, in such a way that quantum spacetimes turn out to be covariant under the action of a suitable quantum kinematical group  of isometries.

Among these noncommutative spacetimes with quantum group symmetry, probably the most relevant example is provided by the well-known $\kappa$-Minkowski noncommutative spacetime \begin{align}
\begin{split}
\label{kminkowski}
&[\hat x^0,\hat x^a] = -\frac{1}{\kappa} \, \hat x^a, \qquad [\hat x^a,\hat x^b] =0, 
\qquad
a,b=1,2,3,
\end{split}
\end{align} 
where $\kappa$ is a parameter proportional to the Planck mass (see~\cite{Maslanka1993,MR1994, Zakrzewski1994poincare, LRNT1991}). The algebra~\eqref{kminkowski} defines a noncommutative spacetime which is covariant under the $\kappa$-Poincar\'e quantum group~\cite{Zakrzewski1994poincare}, a ``quantum deformation" of the group of isometries of Minkowski spacetime which is the (Hopf algebra) dual of the $\kappa$-Poincar\'e quantum algebra, that was obtained for the first time in~\cite{ LRNT1991} (see also~\cite{LNR1991realforms, GKMMK1992, LNR1992fieldtheory}) by making use of quantum group contraction techniques~\cite{CGST1991heisemberg,CGST1992,  BGHOS1995quasiorthogonal} applied onto real forms of the Drinfel'd-Jimbo quantum deformation for appropriate complex simple Lie algebras~\cite{Drinfeld1987icm,Jimbo1985}.
Since then, the $\kappa$-Minkowski spacetime has provided a privileged benchmark for the implementation of  a number of models aiming to describe different features of quantum geometry at the Planck scale and their connections with ongoing phenomenological proposals. Without pretending to be exhaustive, $\kappa$-Minkowski spacetime has been studied in relation with wave propagation on noncommutative spacetimes~\cite{AM2000waves}, Deformed Special Relativity features~\cite{BP2010sigma}, dispersion relations~\cite{BorowiecPachol2009jordanian,BGMP2010,ABP2017}, relative locality phenomena~\cite{ALR2011speed}, curved momentum spaces and phase spaces~\cite{GM2013relativekappa,LSW2015hopfalgebroids}, noncommutative differential calculi~\cite{Sitarz1995plb,JMPS2015}, star products~\cite{DS2013jncg}, noncommutative field theory~\cite{DJMTWW2003epjc, FKN2007plb, DJP2014sigma}, representation theory~\cite{Agostini2007jmp,LMMP2018localization}, light cones~\cite{MS2018plblightcone} and noncommutative spaces of worldlines~\cite{BGH2019worldlinesplb}. 

However, when cosmological distances are involved the interplay between gravity and quantum spacetime should take into consideration the spacetime curvature~\cite{MABGMM2010, AMMR2012, RAMM2015frw, ABDLR2016icecube}, and therefore a natural (maximally symmetric) noncommutative spacetime to be considered should be the quantum analogue of the (Anti-)de Sitter spacetime (hereafter (A)dS). Despite all the efforts devoted so far in the literature to $\kappa$-deformations, the generalization of the $\kappa$-Minkowski spacetime to the (A)dS case with non-vanishing cosmological constant $\Lambda$ was still lacking. The aim of this paper is to fill this gap.

Firstly, we will present the Poisson version of such $\kappa$-(A)dS noncommutative spacetime, since this ``semiclassical" approach to noncommutative spacetimes has been shown to be very efficient from both the conceptual and computational viewpoints (see~\cite{BGH2019worldlinesplb,BMN2017homogeneous}). Secondly, the quantization of the Poisson $\kappa$-(A)dS spacetime will be performed. Throughout this construction the mathematical complexity associated with quantum (A)dS groups will become evident, a fact that should reflect the intertwined features of quantum gravity effects in the case of a quantum spacetime with non-vanishing cosmological constant. 

As a main result, we will show that the quantum $\kappa$-(A)dS spacetime can be consistently and explicitly defined, and turns out to be a cosmological constant deformation of the $\kappa$-Minkowski spacetime~\eqref{kminkowski} in terms of the $\r=\sqrt{-\Lambda}$ parameter. Moreover, we will show that under the physical assumption that the time translation generator is primitive at the $\kappa$-(A)dS quantum algebra level, this is the only non-vanishing cosmological constant  generalization of the $\kappa$-Minkowski spacetime. Explicitly, the  first-order expressions of the $\kappa$-(A)dS spacetime in terms of $\r$ read
\be
\begin{aligned}
&[\hat x^0,\hat x^a] =- \frac{1}{\kappa} \, \hat x^a , \qquad
a=1,2,3,  \\
& [\hat x^1,\hat x^2] =-\,\frac{\r}{\kappa}\,(\hat x^3)^2, \qquad
[\hat x^1,\hat x^3] =\frac{\r}{\kappa}\,\hat x^3\,\hat x^2,\qquad
[\hat x^2,\hat x^3] =-\frac{\r}{\kappa}\,\hat x^1\,\hat x^3 ,
\end{aligned} 
\label{ncstime}
\ee
and the same type of commutation rules will be reproduced when ambient coordinates $\hat s^a$ containing all orders in $\r$ are considered. The most relevant features of this result are:
\begin{itemize}

\item This new noncommutative spacetime is, by construction, covariant under the action of the $\kappa$-(A)dS quantum group in (3+1) dimensions given by the Hopf algebra dual to the $\kappa$-(A)dS quantum algebra  recently presented in~\cite{BHMN2017kappa3+1}.

\item In contradistinction to the $\kappa$-Minkowski case, space coordinates $\hat x^a$ do not commute among themselves and close a homogeneous quadratic algebra~\eqref{ncstime}. As we will see, this algebra can be shown to define a quantum sphere related to the quantum ${\rm SU}(2)\simeq {\rm SO(3)}$ subalgebra of the (3+1)-dimensional $\kappa$-(A)dS quantum group (see also~\cite{BHMN2017kappa3+1}).

\item The $\kappa$-(A)dS spacetime~\eqref{ncstime}  is a smooth deformation of the $\kappa$-Minkowski spacetime~\eqref{kminkowski} in terms of the cosmological constant parameter $\r$. In particular, the $\r\to 0$ limit of all the results contained in this paper will be always well-defined and leads to the corresponding $\kappa$-Minkowski expressions in a straightforward and transparent manner.

\end{itemize}

The structure of the paper is the following. In the next Section we introduce the notion of Poisson homogeneous spaces and their role as semiclassical noncommutative spaces, which we make it explicit for the $\kappa$-Minkowski case and its twisted version.  In Section 3 we review the (A)dS algebra in the kinematical basis. We introduce ambient space coordinates and derive a suitable local coordinate parametrization for the corresponding (A)dS Lie group.  In Section 4 the uniqueness of the proposed generalization of the $\kappa$-deformation to the (A)dS case is rigorously proven, by showing that there only exists  one quantum deformation of the (A)dS algebra that keeps the  time translation generator   primitive and leads to the $\kappa$-Poincar\'e quantum algebra in the vanishing cosmological constant limit. Section 5 presents the full expressions for the semiclassical $\kappa$-(A)dS noncommutative spacetime, and its power series expansion in terms of the cosmological constant parameter $\r$ is analysed. The quantization of this Poisson spacetime is also obtained in local coordinates for the first-order deformation in $\r$, thus giving rise to the quantum $\kappa$-(A)dS spacetime~\eqref{ncstime}. Furthermore the full quantization (in all orders in $\r$)  is performed in ambient coordinates, thus leading to a quadratic algebra whose Casimir operator defines a quantum (A)dS pseudosphere.   A final Section including some comments and open problems closes the paper.


\section{From Poisson to quantum homogeneous spacetimes}

We recall that a Poisson-Lie structure on a Lie group $G$ is a Poisson structure $\{ \cdot, \cdot \} : \mathcal C ^\infty (G) \times \mathcal C ^\infty (G) \rightarrow \mathcal C ^\infty (G) $, that is compatible with the group multiplication $\mu : G \times G \rightarrow G$ in the sense that $\mu$ is a Poisson map. 
In the same way as a Lie algebra is the local counterpart of a Lie group, a Poisson-Lie group has a local structure given by a Lie bialgebra, which is a pair $(\mathfrak g, \delta)$, where $\mathfrak g = \text{Lie} (G)$ and $\delta : \mathfrak g \rightarrow \mathfrak g \wedge \mathfrak g$ is the cocommutator. Note that $\delta$ defines a Lie algebra structure $[\cdot,\cdot]^* : \mathfrak g^* \times \mathfrak g^* \rightarrow \mathfrak g^*$ on the dual vector space $\mathfrak g^*$ of $\mathfrak g$, because the cocycle condition for $\delta$ is equivalent to the Jacobi identity for the dual Lie bracket $[\cdot,\cdot]^*$ (see \cite{ChariPressley1994} for details). 

In particular, for all semisimple Lie algebras all Lie bialgebra structures are coboundary ones, i.e. the cocommutator is   given by 
\begin{align}
\begin{split}
\label{eq:coboundary}
& \delta(X) = [X \otimes 1 + 1 \otimes X,r], \qquad \forall X \in \mathfrak g \, ,
\end{split}
\end{align} 
where the $r$-matrix is a skewsymmetric solution of the modified classical Yang-Baxter equation (mCYBE) on $\mathfrak g$, namely 
\begin{align}
\begin{split}
\label{eq:mCYBE}
&[X \otimes 1 \otimes 1 + 1 \otimes X \otimes 1 + 1 \otimes 1 \otimes X, [[r,r]]\;]=0, \qquad \forall X \in \mathfrak g \,  .
\end{split}
\end{align} 
Here $[[r,r]]$ is the so-called Schouten bracket defined by 
\begin{align}
\begin{split}
\label{eq:schouten}
[[r,r]]=[r_{12},r_{13}]+[r_{12},r_{23}]+[r_{13},r_{23}] \, ,
\end{split}
\end{align} 
where 
$
r_{12}=r^{ij} X_i \otimes X_j \otimes 1, \,  r_{13}=r^{ij} X_i \otimes 1 \otimes X_j$,
$ 
r_{23}=r^{ij} 1 \otimes X_i \otimes X_j
$ and hereafter sum over repeated indices will be assumed.
Coboundary Lie bialgebras are the tangent counterpart of coboundary Poisson-Lie groups, and in that case the unique Poisson-Lie structure on $G$ is given by the so-called Sklyanin bracket,
\begin{align}
\begin{split}
\label{eq:sklyanin}
&\{f,g\}=r^{ij}\left( X^L_i f X^L_j g - X^R_i f X^R_j g\right),\qquad  f,g \in \mathcal C^\infty (G),
\end{split}
\end{align} 
such that $X^L_i$ and $X^R_i$ are left- and right-invariant vector fields defined by
\begin{align}
\label{eq:ivf}
X^L_i f(h)&=\frac{\rm d}{{\rm d} t}\biggr\rvert _{t=0} f\left(h\, {\rm e}^{t T_i}\right), & X^R_i f(h)=\frac{\rm d}{{\rm d}t}\biggr\rvert _{t=0} f\left({\rm e}^{t T_i} h\right) \, ,
\end{align}
where $f \in \mathcal C^\infty(G)$, $h \in G$ and $T_i \in \mathfrak g$.

In this paper we will be interested in studying quotients of Poisson-Lie groups, obtaining in this way covariant Poisson homogeneous spacetimes. In principle, a natural construction would be to consider $G/H$, where $H$ is a Poisson-Lie subgroup of $G$. However, this condition turns out to be too restrictive and it is not necessary in order to have a well-defined Poisson homogeneous space onto $G/H$, as proved in \cite{Drinfeld1993} (see~\cite{BMN2017homogeneous} for a detailed discussion on the subject). In fact, the necessary condition can be stated at the Lie bialgebra level, and is  the so-called coisotropy condition for the cocommutator $\delta$ with respect to the Lie subalgebra $\mathfrak h = \text{Lie}(H)$, namely
\begin{align}
\begin{split}
\label{eq:coisotropycondition}
&\delta(\mathfrak h) \subset \mathfrak h \wedge \mathfrak g \, .
\end{split}
\end{align} 
When this condition is fulfilled for a given $\delta$ and $\mathfrak h$, it can be shown~\cite{BMN2017homogeneous} that the Poisson homogeneous structure on $G/H$ is just defined by the canonical projection of the Sklyanin bracket~\eqref{eq:sklyanin}, and thus can straightforwardly be  obtained provided that a suitable parameterization of the coset space $G/H$ is given. Afterwards, the (comodule algebra) quantization of this Poisson bracket will give rise to the quantum noncommutative space which will be, by construction, covariant under the quantum group defined by the (Hopf algebra) quantization of the Sklyanin bracket. 

In particular, we will be interested in constructing coisotropic Poisson homogeneous spaces for the (3+1)-dimensional Minkowski, dS and AdS spacetimes as homogeneous spacetimes $G/H$, where $G$ is, respectively, the Poincar\'e, ${\rm SO}(4,1)$ and ${\rm SO}(3,2)$ Lie group, and $H={\rm SO}(3,1)$ is always the Lorentz subgroup. In the kinematical basis 
$\{P_0,P_a, K_a, J_a\}$  $(a=1,2,3)$ of generators of time translation, space translations, boosts and rotations,
the commutation rules for the corresponding three Lie algebras can simultaneously be  written in terms of the cosmological constant $\Lambda$ as 
\bea
\begin{array}{lll}
[J_a,J_b]=\epsilon_{abc}J_c ,& \quad [J_a,P_b]=\epsilon_{abc}P_c , &\quad
[J_a,K_b]=\epsilon_{abc}K_c , \\[2pt]
\displaystyle{
  [K_a,P_0]=P_a  } , &\quad\displaystyle{[K_a,P_b]=\delta_{ab} P_0} ,    &\quad\displaystyle{[K_a,K_b]=-\epsilon_{abc} J_c} , 
\\[2pt][P_0,P_a]=-\L  \,K_a , &\quad   [P_a,P_b]=\L \,\epsilon_{abc}J_c , &\quad[P_0,J_a]=0  ,
\end{array}
\label{ads_Liealg3+1}
\eea
where   from now on $a,b,c=1,2,3$. This one-parameter family of Lie algebras contains the dS    algebra $\mathfrak{so}(4,1)$  for   $\L>0$,
the    AdS    algebra $\mathfrak{so}(3,2)$  when  $\L<0$,    and the  Poincar\'e algebra  $\mathfrak{iso}(3,1)$ for  $\L=0$. We will refer to the family of Lie algebras~\eqref{ads_Liealg3+1} as the (A)dS Lie algebra.

In the case of the well-known $\kappa$-Minkowski spacetime, the skewsymmetric solution of the mCYBE (\ref{eq:mCYBE})  which defines the $\kappa$-deformation of the Poincar\'e algebra ($\L=0$)  reads~\cite{Maslanka1993,Zakrzewski1994poincare}
\be
\label{eq:rpoincare}
r_0= \frac{1}{\kappa} \left( K_1 \wedge P_1 + K_2 \wedge P_2 + K_3 \wedge P_3 \right) \, ,
\ee
which generates a quasi-triangular Lie bialgebra with cocommutator map given by
\begin{align}
\begin{split}
\label{eq:deltakappaPoincare}
&\delta(P_0)=\delta(J_a)=0 ,\\
&\delta(P_a)= \frac{1}{\kappa} P_a \wedge P_0 ,\\
&\delta(K_a)= \frac{1}{\kappa} (  K_a \wedge P_0+ \epsilon_{abc} P_b \wedge J_c) .
\end{split}
\end{align} 
This Lie bialgebra defines a coisotropic Poisson homogeneous Minkowski spacetime  $G/H$ since the coisotropy  condition~\eqref{eq:coisotropycondition} for the Lorentz subalgebra $\mathfrak h = \{ K_a,J_a\}$ is fulfilled. We remark that despite $H$ is not endowed with Poisson-Lie subgroup structure, all coisotropic deformations generate well-defined Poisson homogeneous spaces. 

Therefore, the $\kappa$-Minkowski Poisson homogeneous spacetime can be obtained by constructing the appropriate parametrization of the Poincar\'e group such that $(x^0,x^1,x^2,x^3)$ provide suitable coordinates for $G/H$, and then computing the Sklyanin bracket~\eqref{eq:sklyanin} for such Minkowski subalgebra. This explicit construction (see~\cite{BGH2019worldlinesplb} for details) reads 
\begin{align}
\begin{split}
\label{eq:PoissonMinkowski}
&\{x^0,x^a\} = -\frac{1}{\kappa}\, x^a, \qquad \{x^a,x^b\} =0 .
\end{split}
\end{align} 
From these expressions the quantization giving rise to the well-known $\kappa$-Minkowski spacetime~\eqref{kminkowski} is straightforward, since the algebra~\eqref{eq:PoissonMinkowski} is linear. Also, the Sklyanin bracket is such that rapidities and coordinates for rotation angles Poisson-commute, and the complete quantum $\kappa$-Poincar\'e group can be obtained~\cite{Maslanka1993}.

This approach can be applied to any other coisotropic deformation. For instance, the $r$-matrix generating a  twisted version of the $\kappa$-Poincar\'e algebra~\cite{Daszkiewicz2008} 
\be
\label{eq:rpoincaretwist}
\tilde r_0= r_0 + r^t= \frac{1}{\kappa} \left( K_1 \wedge P_1 + K_2 \wedge P_2 + K_3 \wedge P_3 \right) + \t J_3  \wedge  P_0 \, ,
\ee
has an additional twist term $r^t$ which preserves the coisotropy condition for the Lorentz subalgebra $\mathfrak h$. Therefore, the projection of the corresponding Sklyanin bracket to the Minkowski space provides a twisted $\kappa$-Minkowski spacetime, namely
\begin{align}
\begin{split}
\label{eq:PoissonMinkowski_twist}
&\{x^0,x^1\} = -\frac{1}{\kappa}  \,x^1 - \t \,x^2, \qquad \{x^0,x^2\} = -\frac{1}{\kappa} \,x^2 + \t \, x^1, \qquad \{x^0,x^3\} = -\frac{1}{\kappa}  \,x^3, \\
& \{x^a,x^b\} =0,  
\end{split}
\end{align} 
 whose quantization is also straightforward.


\section{Local coordinates for (A)dS spacetimes}

According to the previous approach and
with the aim to construct the  $\kappa$-(A)dS noncommutative spacetime, we need a suitable parametrization of the (A)dS group and the corresponding       Lorentzian    homogeneous
space $G/H$. 
 Let us consider the vector representation of the  (A)dS  Lie algebra (\ref{ads_Liealg3+1}), $\rho : \mathfrak{g} \rightarrow \text{End}(\mathbb R ^5)$, where a generic Lie algebra element $X$ reads
\begin{equation}
\label{eq:repG}
\rho(X)=  x^0 \rho(P_0)  +  x^a \rho(P_a)  +  \xi^a \rho(K_a) +  \theta^a \rho(J_a) =\left(\begin{array}{ccccc}
0&\L\,x^0&-\L\,x^1&-\L\,x^2&-\L\,x^3\cr 
x^0 &0&\xi^1&\xi^2&\xi^3\cr 
x^1 &\xi^1&0&-\theta^3&\theta^2\cr 
x^2 &\xi^2&\theta^3&0&-\theta^1\cr 
x^3 &\xi^3&-\theta^2&\theta^1&0
\end{array}\right) .
\end{equation}
This faithful representation $\rho$ can be exponentiated to the following  (A)dS group element  by using the local coordinates $(x^0,\boldsymbol{x},\boldsymbol{\xi},\boldsymbol{\theta})$  (where hereafter we denote $\boldsymbol{y}=(y^1,y^2,y^3)$):
\begin{align}
\begin{split}
\label{eq:Gm}
&G_\L= \exp{x^0 \rho(P_0)} \exp{x^1 \rho(P_1)} \exp{x^2 \rho(P_2)} \exp{x^3 \rho(P_3)} \\
&\qquad\quad\times \exp{\xi^1 \rho(K_1)} \exp{\xi^2 \rho(K_2)} \exp{\xi^3 \rho(K_3)}
 \exp{\theta^1 \rho(J_1)} \exp{\theta^2 \rho(J_2)} \exp{\theta^3 \rho(J_3)} .
\end{split}
\end{align}
Note that the Lorentz subgroup $H= {\rm SO} (3,1)$ is parametrized by the six exponentials containing the generators $\mathfrak h = \{ K_a,J_a\}$. The previous ordering guarantees that the spacetime coordinates $(x^0,\boldsymbol{x})$ can be interpreted as the right coset coordinates for any value of the cosmological constant $\L$, since they come from the factorization of  $G_{\L}=T_\L \cdot H$ where $T_\L$ and $H$ are, respectively, the translations and Lorentz sectors.

The matrix representation~\eqref{eq:repG} allows us to identify the (A)dS  Lie group as  the isometry group  of the 5-dimensional linear space  $(\mathbb R^5, \mathbf I_{\L})$ with ambient  coordinates
$ (\s^4,\s^0,\s^1,\s^2,\s^3)\equiv  (\s^4,\s^0,\boldsymbol{\s})$ such that $\mathbf I_{\L}$ is the bilinear form given by
\begin{align}
&\mathbf I_{\L}={\rm diag}(+1,-\L,\L,\L ,\L),
\label{bf}
\end{align}
and ({\ref{eq:Gm}) fulfils $ G_\L^T \, \mathbf I_{\L}\, G_\L =\mathbf I_{\L} $.  In this way the origin of the (A)dS   spacetime has ambient coordinates $O =(1,0,0,0,0)$ and   is invariant under the action of the Lorentz subgroup $ H$ (see~\eqref{eq:repG}). The orbit
passing through $O$ corresponds to the (3+1)-dimensional  (A)dS spacetime  defined by the pseudosphere
\begin{equation}
\Sigma_\L\equiv ( s^4)^2 - \L  (\s^0)^2 +\L \bigl( (\s^1)^2+ (\s^2)^2+ (\s^3)^2 \bigr)=1  ,
\label{pseudo}
\end{equation}
determined by $\mathbf I_{\L}$ (\ref{bf}).  Note that   in the limit   $\L\to 0$, the Minkowski spacetime will be identified with the hyperplane $\s^4=+1$, which also contains $O$.

The spacetime coordinates $(x^0,\boldsymbol{x})$ are the so-called {geodesic
parallel coordinates} (see~\cite{BHMN2014sigma}) which are defined in terms of the action of the one-parameter subgroups of spacetime translations onto the origin $O=(1,0,0,0,0)$ through
\be
(\s^4,\s^0,\>\s )^T=\exp{x^0 \rho(P_0)} \exp{x^1 \rho(P_1)} \exp{x^2 \rho(P_2)} \exp{x^3 \rho(P_3)}\,\cdot O^T .
\ee
This yields the following relationships between ambient and local coordinates for  the (A)dS   spacetime:
\begin{align} 
\begin{split}
\label{ambientspacecoords}
&\s^4=\cos \r x^0 \cosh \r x^1 \cosh \r x^2\cosh \r x^3 , \\
&\s^0=\frac {\sin \r x^0}\r  \cosh \r x^1 \cosh \r x^2\cosh \r x^3, \\
&\s^1=\frac {\sinh \r x^1 }\r   \cosh \r x^2\cosh \r x^3, \\
&  \s^2=\frac { \sinh \r x^2} \r\cosh \r x^3, \\
&  \s^3=\frac { \sinh \r x^3} \r ,
\end{split}
\end{align}
where the parameter  $\ro$ is defined by 
\be
\ro^2:=-\Lambda .
\label{constant}
\ee
Thus $\ro$ is real for the AdS space
and a purely imaginary number    for the dS one. In the vanishing cosmological limit, $\ro\to 0$, the ambient coordinates reduce to the usual Cartesian ones in the Minkowski spacetime: $(\s^4,\s^0,\>\s )\equiv (1,x^0,\>x )$.
 The  metric on the homogeneous  spacetime can  now be  obtained 
from the flat ambient metric determined by $\mathbf I_{\L}$, after dividing it  by the curvature (which is $-\L$) and
by restricting the resulting metric to the pseudosphere $\Sigma_\L$ (\ref{pseudo}). Finally, its expression  in terms of geodesic parallel coordinates turns out to be
\bea
&&\dd\sigma^2 =\cosh^2(\r x^1) \cosh^2(\r x^2)\cosh^2(\r x^3) (\dd x^0)^2-\cosh^2(\r
x^2) \cosh^2(\r x^3)(\dd x^1)^2 \nonumber\\ 
&&\qquad \quad    -\cosh^2(\r x^3)( \dd x^2)^2- (\dd x^3)^2 \, .
\eea

Note also that the explicit form of the (A)dS group element   $G_\L$ (\ref{eq:Gm}) reads
\be
G_\L=\left(\begin{array}{ccccc}
\s^4&\aaa^4_{0}&\aaa^4_{1}&\aaa^4_{2}&\aaa^4_{3}\\[2pt] 
\s^0&\bbb^0_{0}&\bbb^0_{1}&\bbb^0_{2}&\bbb^0_{3}\\[2pt] 
\s^1&\bbb^1_{0}&\bbb^1_{1}&\bbb^1_{2}&\bbb^1_{3}\\[2pt] 
\s^2&\bbb^2_{0}&\bbb^2_{1}&\bbb^2_{2}&\bbb^2_{3}\\[2pt] 
\s^3&\bbb^3_{0}&\bbb^3_{1}&\bbb^3_{2}&\bbb^3_{3}
\end{array}\right) ,  
\label{groupw}
\ee
where the entries $\aaa^\alpha_{\beta}$ and $\bbb^\mu_{\nu}$ depend on all the group coordinates $(x^0,\>x,\boldsymbol{\xi},\boldsymbol{\theta})$ and on  the cosmological constant   $\L$. Recall that   translations do not close a subgroup and that the action of  the  (A)dS group on the coordinates is not linear. In the limit $\L\to 0$, these expressions reduce to the well-known matrix representation of the  Poincar\'e group  
\be
\lim_{\L \to 0} G_\L=\left(\begin{array}{ccccc}
1&0&0&0&0\cr 
x^0&L^0_{0}&L^0_{1}&L^0_{2}&L^0_{3}\\[2pt]  
x^1&L^1_{0}&L^1_{1}&L^1_{2}&L^1_{3}\\[2pt]  
x^2&L^2_{0}&L^2_{1}&L^2_{2}&L^2_{3}\\[2pt]  
x^3&L^3_{0}&L^3_{1}&L^3_{2}&L^3_{3}
\end{array}\right) ,  
\ee
such that the entries $L^\mu_{\nu}$ parametrize an element of the Lorentz subgroup, so depending only on $(\boldsymbol{\xi},\boldsymbol{\theta})$.
From the group action of $G_\L$~\eqref{groupw} on itself  via right- and left-multiplication, a lengthy computation provides explicit expressions for  left- and right-invariant vector fields $X^L_i$ and $X^R_i$ (\ref{eq:ivf}) in terms of the local coordinates $(x^0,\>x,\boldsymbol{\xi},\theta)$ and $\L$.

   
\section{$\kappa$-deformations of the (A)dS algebra}

As it was previously mentioned, the $\kappa$-Poincar\'e deformation is completely determined by the solution of  the mCYBE (\ref{eq:mCYBE})  given by the skewsymmetric element~\eqref{eq:rpoincare}. The main feature of this $r$-matrix is that the associated cocommutator $\delta(P_0)$ vanishes, and this is a necessary condition for the coproduct of the $\kappa$-Poincar\'e quantum algebra   to be a primitive generator, namely $\Delta(P_0)=P_0 \otimes 1+1\otimes P_0 $.
This fact is essential in order to allow exponentials ${\rm e}^{P_0/\kappa}$ to emerge as the building blocks of the quantum $\kappa$-deformation and of the dispersion relation arising from the deformed Casimir, thus implying that $\kappa$ has dimensions of a (Planck) mass. 

Therefore, it seems natural to assume that  a quantum deformation of the (A)dS algebra can properly be  called a $\kappa$-deformation provided that it is generated by a skewsymmetric solution $r_\L$ of the mCYBE for the (A)dS algebra (\ref{ads_Liealg3+1}) fulfilling two conditions:
\begin{itemize}
\item The $P_0$ generator is primitive: $\delta(P_0)=[P_0 \otimes 1 + 1 \otimes P_0,r_\L]=0$.
\item Its vanishing cosmological constant limit is just the $\kappa$-Poincar\'e $r$-matrix~\eqref{eq:rpoincare}, namely
$
\lim_{\L \rightarrow 0} r_\L = r_0.
$
\end{itemize}

A long but straightforward computer-assisted calculus (which starts from a completely generic skewsymmetric $r$-matrix depending on 45 parameters onto which the mCYBE is imposed) shows that the only family of multiparametric (A)dS $r$-matrices compatible with these two conditions is given by:
\begin{align}
\begin{split}
\label{eq:r_firstfamily}
&r_\L=\frac{1}{\kappa}( K_1 \wedge P_1 + K_2 \wedge P_2 + K_3 \wedge P_3 ) + P_0 \wedge (\p_1 J_1 + \p_2 J_2 + \p_3 J_3) \\
&\qquad\qquad+ \pp_3 J_1 \wedge J_2 - \pp_2 J_1 \wedge J_3 + \pp_1 J_2 \wedge J_3 \, ,
\end{split}
\end{align} 
together with the following quadratic relations among the parameters:
\begin{align}
\begin{split}
\label{eq:mCYBE_firstfamily}
\p_1 \pp_3 - \p_3 \pp_1 = 0, \qquad &\p_1 \pp_2 - \p_2 \pp_1 =0, \qquad \p_2 \pp_3 - \p_3 \pp_2 = 0, \\
&\pp_1^2 + \pp_2^2 + \pp_3^2 = \left( \frac{\r}{\kappa} \right)^2  .
\end{split}
\end{align} 
Notice that the term $P_0 \wedge (\p_1 J_1 + \p_2 J_2 + \p_2 J_3)$ in \eqref{eq:r_firstfamily} is given by the superposition of three twists (recall from~\eqref{ads_Liealg3+1} that $[P_0,J_a]=0$) and therefore these three terms would lead to the (A)dS generalization of the twisted $\kappa$-Poincar\'e $r$-matrix~\eqref{eq:rpoincaretwist}. 
The  equations (\ref{eq:mCYBE_firstfamily})  have a neat geometrical interpretation: non-twisted solutions (with parameters $\pp_i$) are given by the vector of a point in the sphere with radius $\r/\kappa$, while twisted solutions (with parameters $\p_i$)  are defined by another vector orthogonal to the former. Note also that equations~\eqref{eq:mCYBE_firstfamily} are valid for $\L = 0$ ($\r=0$); in this Poincar\'e case $\pp_1 = \pp_2 = \pp_3 = 0$ and the twists parameters are free.

In order to solve the equations \eqref{eq:mCYBE_firstfamily}, let us firstly consider  the non-twisted case with $\p_1 = \p_2 = \p_3 = 0$. Then the only non-vanishing equation in~\eqref{eq:mCYBE_firstfamily} defines a  sphere of radius $R= \r / \kappa$, so we can write 
\be
\label{eq:parametrizationsphere}
\pp_3 =R \cos \theta, \qquad
\pp_2 = -R \sin \theta \sin \varphi, \qquad
\pp_1 = R \sin \theta \cos \varphi,
\ee
where $\theta \in [0,\pi], \varphi \in [0,2 \pi)$. Now, the solution~\eqref{eq:r_firstfamily} reads
\begin{align}
\begin{split}
\label{eq:r_firstfamily_notwist}
&r_\L=\frac{1}{\kappa}(K_1 \wedge P_1 + K_2 \wedge P_2 + K_3 \wedge P_3)   \\
&\qquad\qquad +\frac{\r}{\kappa} ( \cos \theta J_1 \wedge J_2 + \sin \theta \sin \varphi J_1 \wedge J_3 + \sin \theta \cos \varphi J_2 \wedge J_3) \, .
\end{split}
\end{align} 
The last term within the $r$-matrix~\eqref{eq:r_firstfamily_notwist} is represented by a point on the 2D sphere parametrized by \eqref{eq:parametrizationsphere}, and it is straightforward to prove that the Lie algebra generator 
\begin{align}
\begin{split}
\tilde J_3 = \sin \theta \cos \varphi J_1 - \sin \theta \sin \varphi J_2 + \cos \theta J_3 \, ,
\end{split}
\label{vectorJ}
\end{align} 
becomes primitive under the deformation defined by the $r$-matrix~\eqref{eq:r_firstfamily_notwist}, i.e $\delta(\tilde J_3)=0$. Now, since there exists an  
     automorphism of the (A)dS algebra~\eqref{ads_Liealg3+1} 
  that corresponds to the rotation providing the    new $\tilde J_3$ generator~\eqref{vectorJ}, we can apply it to the $r$-matrix (\ref{eq:r_firstfamily_notwist}), and we find the following  transformed $r$-matrix (tildes will be omitted for the sake of simplicity)
\begin{align}
\begin{split}
\label{eq:r_firstfamily_notwist_j3}
&r_\L=\frac{1}{\kappa}( K_1 \wedge P_1 + K_2 \wedge P_2 + K_3 \wedge P_3 + \r  J_1 \wedge  J_2)\, .
\end{split}
\end{align} 
This shows that we can simply take $\theta = 0$ in~\eqref{eq:r_firstfamily_notwist} with no loss of generality, and we arrive at the only possible solution for the $r$-matrix which has previously been  considered as the one generating the (non-twisted) $\kappa$-(A)dS deformation~\cite{BGHOS1995quasiorthogonal,BHMN2017kappa3+1,BGGH2017curvedplb,BGGH2018cms31}. Moreover, this computation provides a neat geometrical intuition of the fact discussed in \cite{BHM2014tallinn} that a rotation generator becomes privileged when $\L\neq 0$.  
Also, this proves that, modulo Lie algebra automorphisms,  the (A)dS  $r$-matrix~\eqref{eq:r_firstfamily_notwist_j3} is the only (non-twisted) skewsymmetric solution of the mCYBE which generalizes the $\kappa$-Poincar\'e deformation.

For the twisted case we have that $(\p_1, \p_2, \p_3) \not = (0,0,0)$. With no loss of generality we can assume that $\p_3 \neq 0$. By taking into account (\ref{eq:mCYBE_firstfamily}) and (\ref{eq:parametrizationsphere})  we find that 
\begin{equation}
\p_1= \p_3 \tan \theta \cos \varphi, \qquad
\p_2=-\p_3 \tan \theta \sin \varphi,
\end{equation}
($\theta \not = \pi / 2$) which  inserted in \eqref{eq:r_firstfamily_notwist} gives 
\begin{align}
\begin{split}
\label{eq:r_firstfamily_twist}
&r_\L=\frac{1}{\kappa}(K_1 \wedge P_1 + K_2 \wedge P_2 + K_3 \wedge P_3 ) + \p_3 P_0 \wedge ( \tan \theta \cos \varphi J_1 - \tan \theta \sin \varphi J_2 +  J_3)  \\
&\qquad\qquad + \frac{\r}{\kappa}  ( \cos \theta J_1 \wedge J_2 + \sin \theta \sin \varphi J_1 \wedge J_3 + \sin \theta \cos \varphi J_2 \wedge J_3) .
\end{split}
\end{align} 
Now, if we consider the rotated basis such that $\theta = 0$ and rename the twist parameter as $\p_3 =- \t$ we arrive at
\begin{align}
\begin{split}
\label{eq:r_firstfamily_twist_j3}
&r_\L=\frac{1}{\kappa}( K_1 \wedge P_1 + K_2 \wedge P_2 + K_3 \wedge P_3 + \r  J_1 \wedge  J_2) + \t  J_3   \wedge P_0\, ,
\end{split}
\end{align} 
which is just the $r$-matrix presented in~\cite{BHN2015towards31} as the one arising from a Drinfel'd double structure of the (A)dS Lie algebra (see also~\cite{BHMN2017kappa3+1}). The Poincar\'e $\L\to 0$  limit of this $r$-matrix  yields (\ref{eq:rpoincaretwist}), which along with   its Galilean counterpart were studied in~\cite{Daszkiewicz2008}.


\section{The $\kappa$-{(A)dS} noncommutative spacetime}

After the previous discussion we shall take the $r$-matrix~\eqref{eq:r_firstfamily_notwist_j3} as the generating object for the $\kappa$-{(A)dS}  deformation, and in particular for its associated noncommutative spacetime. First of all, we compute the cocommutator map (\ref{eq:coboundary}) associated to this $r$-matrix, which reads 
\begin{align}
\begin{split}
\label{eq:deltakappaAdS}
& \delta(P_0)=\delta(J_3)= 0, \qquad \delta(J_1)=\frac \r\kappa J_1 \wedge J_3, \qquad \delta(J_2)= \frac \r\kappa   J_2 \wedge J_3 ,\\
& \delta(P_1)= \frac{1}{\kappa} (P_1 \wedge P_0 - \r P_3 \wedge J_1 - \r^2 K_2 \wedge J_3 + \r^2 K_3 \wedge J_2) ,\\
& \delta(P_2)= \frac{1}{\kappa} (P_2 \wedge P_0 - \r P_3 \wedge J_2 + \r^2 K_1 \wedge J_3 - \r^2 K_3 \wedge J_1), \\
& \delta(P_3)= \frac{1}{\kappa} (P_3 \wedge P_0 + \r P_1 \wedge J_1 + \r P_2 \wedge J_2 - \r^2 K_1 \wedge J_2 + \r^2 K_2 \wedge J_1), \\
& \delta(K_1)= \frac{1}{\kappa} (K_1  \wedge P_0  + P_2 \wedge J_3 - P_3 \wedge J_2 - \r K_3 \wedge J_1) ,\\
& \delta(K_2)= \frac{1}{\kappa} ( K_2 \wedge P_0  - P_1 \wedge J_3 + P_3 \wedge J_1 - \r K_3 \wedge J_2) ,\\
& \delta(K_3)= \frac{1}{\kappa} ( K_3 \wedge P_0  + P_1 \wedge J_2 - P_2 \wedge J_1 + \r K_1 \wedge J_1 + \r K_2 \wedge J_2).
\end{split}
\end{align} 
Here it becomes clear that the $\mathfrak{su}(2)\simeq \mathfrak{so}(3)$ Lie subalgebra generated by the rotation generators $\{J_1,J_2,J_3\}$ defines a sub-Lie bialgebra structure, which becomes non-trivial when the cosmological constant \eqref{constant} is different from zero, a fact that will be relevant in the sequel.

Since the cocommutator~\eqref{eq:deltakappaAdS} fulfils the coisotropy condition  (\ref{eq:coisotropycondition}) with respect to the Lorentz subalgebra, the Poisson homogeneous $\kappa$-(A)dS spacetime will be given by the Sklyanin bracket~\eqref{eq:sklyanin} restricted to the spacetime coordinates. Threfore, left- and right-invariant vector fields for the (A)dS group have to be obtained (through a really cumbersome computer-assisted computation) from the group action given by the multiplication of two matrices~\eqref{groupw}. For the sake of brevity we omit the explicit form of such invariant vector fields, which after being inserted in the Sklyanin bracket defined by~\eqref{eq:r_firstfamily_notwist_j3} lead to the  Poisson homogeneous $\kappa$-(A)dS spacetime given by
\begin{align}
\begin{split}
\label{eq:Poisson-kappaAdSst}
&\{x^0,x^1\} =-\frac{1}{\kappa}\, \frac{\tanh (\r x^1)}{\r \cosh^2(\r x^2) \cosh^2(\r x^3)} ,\\
&\{x^0,x^2\} =-\frac{1}{\kappa}\,\frac{\tanh (\r x^2)}{\r \cosh^2(\r x^3)} ,\\
&\{x^0,x^3\} =-\frac{1}{\kappa}\,\frac{\tanh (\r x^3)}{\r},
\end{split}
\end{align} 
\begin{align}
\begin{split}
\label{eq:Poisson-kappaAdSss}
&\{x^1,x^2\} =-\frac{1}{\kappa}\,\frac{\cosh (\r x^1) \tanh ^2(\r x^3)}{\r} ,\\
&\{x^1,x^3\} =\frac{1}{\kappa}\,\frac{\cosh (\r x^1) \tanh (\r x^2) \tanh (\r x^3)}{\r} ,\\
&\{x^2,x^3\} =-\frac{1}{\kappa}\,\frac{\sinh (\r x^1) \tanh (\r x^3)}{\r} \, ,
\end{split}
\end{align} 
which can be thought of as a (complicated) cosmological constant deformation of the (Poisson) $\kappa$-Minkowski spacetime~\eqref{eq:PoissonMinkowski} in terms of the parameter $\r$ (\ref{constant}).

A striking feature of the $\kappa$-(A)dS spacetime suddenly arises from these expressions: brackets between space coordinates do not vanish, in contradistinction with the $\kappa$-Minkowski case (and also with the (2+1) $\kappa$-(A)dS spacetime presented in~\cite{ BHMN2014sigma}, which can be obtained from~\eqref{eq:Poisson-kappaAdSst}--\eqref{eq:Poisson-kappaAdSss} by projecting $x^3 \rightarrow 0$). In order to stress the relationship with the $\kappa$-Minkowski expressions, we can take the power series expansion of~\eqref{eq:Poisson-kappaAdSst} in terms of $\r$, and we get
\begin{align}
\begin{split}
\label{eq:Poisson-kappaAdSst1}
&\{x^0,x^1\} =- \frac{1}{\kappa}\, (x^1 + o[\r^2]), \\
&\{x^0,x^2\} =- \frac{1}{\kappa}\, (x^2 + o[\r^2]), \\
&\{x^0,x^3\} =- \frac{1}{\kappa}\, (x^3 + o[\r^2]) ,
\end{split}
\end{align} 
whose zeroth-order in $\eta$ is just the $\kappa$-Minkowski spacetime, whilst the first-order deformation in $\r$ of the space subalgebra~\eqref{eq:Poisson-kappaAdSss} defines the following homogeneous quadratic algebra
\begin{align}
\begin{split}
\label{eq:Poisson-kappaAdSss1}
&\{x^1,x^2\} =-\frac{1}{\kappa}\,(\r\,(x^3)^2 + o[\r^2]), \\
&\{x^1,x^3\} =\frac{1}{\kappa}\,(\r\,x^2 x^3 + o[\r^2]),\\
&\{x^2,x^3\} =-\frac{1}{\kappa}\,(\r\,x^1 x^3 + o[\r^2]).
\end{split}
\end{align} 

This essential novelty of the $\kappa$-(A)dS spacetime deserves further discussion. Firstly, note that the quadratic Poisson algebra arising in~\eqref{eq:Poisson-kappaAdSss1} and given by
\begin{equation}
\{x^1,x^2\} =-\frac{\r}{\kappa}\,\,(x^3)^2, \qquad
\{x^1,x^3\} =\frac{\r}{\kappa}\, x^2 x^3,\qquad
\{x^2,x^3\} =-\frac{\r}{\kappa}\,x^1 x^3 ,
\label{quadraticsu2}
\end{equation}
can be identified~\cite{Grabowski1990brno,Grabowski1995} as a subalgebra of the semiclassical limit of Woronowicz's quantum ${\rm SU}(2)$ group~\cite{Woronowicz1987tsu2,Woronowicz1987cmp} (see also~\cite{VaksmanSoibelman1988,ChaichianDemichev1996book}). We also recall that the brackets
\begin{equation}
\{x^1,x^2\}=f\,\frac{\partial F}{\partial x^3}, \qquad
\{x^2,x^3\}=f\,\frac{\partial F}{\partial x^1}, \qquad
\{x^3,x^1\}=f\,\frac{\partial F}{\partial x^2},
\label{3dpois}
\end{equation}
always define  a three-dimensional Poisson algebra for any choice of the smooth functions $f$ and $F$, and the Casimir function for~\eqref{3dpois} is just the function $F$~\cite{FG1994quadpoisson}. Therefore, the algebra~\eqref{quadraticsu2} can directly be  obtained by taking
\be
F(x^1,x^2,x^3)=(x^1)^2 + (x^2)^2 + (x^3)^2,\qquad
f(x^1,x^2,x^3)=-\frac12\, \frac{\r}{\kappa}\,x^3.
\ee
This implies that two-dimensional spheres
\be
S=(x^1)^2 + (x^2)^2 + (x^3)^2,
\label{sphere}
\ee
define symplectic leaves for the Poisson structure~\eqref{quadraticsu2}. Moreover, it is straightforward to check that the Poisson brackets~\eqref{eq:Poisson-kappaAdSss} arise in the Sklyanin bracket just from the $J_1\wedge J_2$ term of the $r$-matrix~\eqref{eq:r_firstfamily_notwist_j3}. This explains why the Poisson algebra~\eqref{quadraticsu2} is naturally linked to the semiclassical limit of the quantum ${\rm SU}(2)$ subgroup of the $\kappa$-(A)dS deformation, albeit realized on the 3-space coordinates. In this respect, we recall that the $\mathfrak{su}(2)$ subalgebra generated by $\{J_1,J_2,J_3 \}$ becomes a quantum $\mathfrak{su}(2)$ subalgebra when the full quantum deformation is constructed~\cite{BHMN2017kappa3+1}, a fact that can already be  envisaged from the cocommutator~\eqref{eq:deltakappaAdS} where the $\mathfrak{su}(2)$ generators define a sub-Lie bialgebra.

Furthermore, the algebra~\eqref{quadraticsu2} can be quantized as
\be
[\hat x^1,\hat x^2] =-\,\frac{\r}{\kappa}\,(\hat x^3)^2, \qquad
[\hat x^1,\hat x^3] =\,\frac{\r}{\kappa}\,\hat x^3 \hat x^2,\qquad
[\hat x^2,\hat x^3] =-\,\frac{\r}{\kappa}\,\hat x^1\hat x^3 ,
\label{quantumsu2}
\ee
since associativity is ensured by the Jacobi identity, which can  be   checked by considering the ordered monomials $(\hat x^1)^l\,(\hat x^3)^m\,(\hat x^2)^n$. The Casimir operator for~\eqref{quantumsu2} can be proven to be
\be
\hat S_{\eta/\kappa}=(\hat x^1)^2 + (\hat x^2)^2 + (\hat x^3)^2  + \frac{\r}{\kappa}\, \hat x^1 \hat x^2,
\label{qcas}
\ee
which defines the ``quantum spheres" generated by the noncommuting $\kappa$-(A)dS local coordinates. Thus space coordinates become noncommutative, while at first-order in $\r$ the time-space sector is kept invariant with respect to the $\kappa$-Minkowski case. Moreover, the space-space brackets~\eqref{quadraticsu2} are just a subalgebra of the quantum ${\rm SU}(2)$ group.

Now, the quantum $\kappa$-(A)dS spacetime for any order in $\r$ should be obtained as the quantization of the full Poisson algebra~\eqref{eq:Poisson-kappaAdSst}-\eqref{eq:Poisson-kappaAdSss}, which is by no means a trivial task due to the noncommutativity of the space coordinates given by~\eqref{eq:Poisson-kappaAdSss}. However, by considering the five ambient coordinates $(\s^4,\s^0,\>\s )$ defined by \eqref{ambientspacecoords} and fulfilling the constraint (\ref{pseudo}), we get that their Sklyanin bracket leads to the following   quadratic algebra 
\begin{equation}
\label{eq:Poisson-kappaAdSambient}
\begin{aligned}
&  \{s^0,s^a\} = -\frac{1}{\kappa}\, s^{a} s^{4} ,\qquad \{s^4,s^a\} = \frac{\r^{2}}{\kappa}\, s^{a}  s^{0} ,\qquad a=1,2,3,\\[2pt]
 &\{s^1,s^2\} = -\frac{\r}{\kappa} (s^{3})^{2} ,\qquad
 \{s^1,s^3\} = \frac{\r}{\kappa}\, s^{2} s^{3} ,\qquad
 \{s^2,s^3\} = -\frac{\r}{\kappa}\, s^{1} s^{3} ,\\
&\{s^0,s^4\} = -\frac{\r^{2}}{\kappa}  \left((s^{1})^{2} + (s^{2})^{2} +(s^{3})^{2} \right)  ,
\end{aligned}
\end{equation}
which is, at most, quadratic in the cosmological constant parameter $\r$. Since the subalgebra generated by the three ambient space coordinates $\>\s$ is formally the same as~\eqref{quadraticsu2}, its quantization would give the same result as~\eqref{quantumsu2}, but now with $\hat {\>\s}$ instead of $\hat {\>x}$. By taking into account this fact and by   considering the  ordered monomials $(\hat s^0)^k\,(\hat s^1)^l\,(\hat s^3)^m\,(\hat s^2)^n\,(\hat s^4)^j$, a long but straightforward computation shows that 
the following quadratic brackets give rise to an associative algebra  (i.e.,~Jacobi identities  are satisfied) which becomes the full quantization of the Poisson brackets (\ref{eq:Poisson-kappaAdSambient})
\begin{equation}
\label{eq:Poisson-kappaAdSambient2}
\begin{aligned}
&  [\hat s^0, \hat s^a] = -\frac{1}{\kappa}\, \hat s^{a} \hat s^{4} ,\qquad [\hat s^4,\hat s^a] = \frac{\r^{2}}{\kappa}\, \hat s^{0}  \hat s^{a}  ,\qquad [\hat s^0,\hat s^4] = -\frac{\r^{2}}{\kappa} \, \hat \esf_{\r/\kappa} ,\\[2pt]
 &[\hat s^1,\hat s^2] = -\frac{\r}{\kappa} (\hat s^{3})^{2} ,\qquad
 [\hat s^1,\hat s^3] = \frac{\r}{\kappa}\, \hat s^{3}  \hat s^{2} ,\qquad
 [\hat s^2,\hat s^3] = -\frac{\r}{\kappa}\, \hat s^{1} \hat s^{3} ,
 \end{aligned}
\end{equation}
and defines the $\kappa$-(A)dS spacetime for all orders in $\r$. Here $\hat \esf_{\r/\kappa} $ is given by
\be
\hat \esf_{\r/\kappa}= (\hat s^1)^2 + (\hat s^2)^2 + (\hat s^3)^2  + \frac{\r}{\kappa}\, \hat s^1 \hat s^2 \, ,
\label{qcas2}
\ee
and this operator is the analogue of the quantum sphere~\eqref{qcas}  in quantum ambient coordinates, since~\eqref{qcas2} is just the Casimir operator for the subalgebra spanned by   $\hat{\>\s}$, namely $[\hat \esf_{\r/\kappa} , \hat s^a]=0$. However, $\hat \esf_{\r/\kappa} $ does not commute with the remaining 
quantum ambient coordinates:
\be
\begin{aligned}
&[\hat \esf_{\r/\kappa} ,\hat s^0]=\frac 1 \kappa \bigl(\hat s^4\, \hat \esf_{\r/\kappa} + \hat \esf_{\r/\kappa}\,  \hat s^4 \bigr)-\frac{\r^2}{\kappa^2}\,\hat s^0\,  \hat\esf_{\r/\kappa},\\[2pt]
&[\hat \esf_{\r/\kappa} ,\hat s^4]=-\frac {\r^2} \kappa \bigl(\hat s^0\, \hat \esf_{\r/\kappa} + \hat \esf_{\r/\kappa}\,  \hat s^0 \bigr)+\frac{\r^2}{\kappa^2}\,  \hat\esf_{\r/\kappa} \,\hat s^4 .
\end{aligned}
\ee
In fact, the Casimir operator for the full $\kappa$-(A)dS quantum space~(\ref{eq:Poisson-kappaAdSambient2}) is found to be
\be
\hat \Sigma_{\r,\kappa}= (\hat s^4)^2 +\r^2 (\hat s^0)^2 - \frac{\r^2}{\kappa} \, \hat s^0 \hat s^4  - \r^2 \hat \esf_{\r/\kappa} ,
\label{qcas3}
\ee
which is just the quantum analogue of the pseudosphere (\ref{pseudo}) that defines the (A)dS space.

Indeed, the quantization of the algebra~\eqref{eq:Poisson-kappaAdSambient} that we have obtained should coincide with the corresponding subalgebra of the full $\kappa$-(A)dS quantum group relations obtained by applying the usual FRT approach~\cite{RTF1990} onto the quantum matrix group arising from~\eqref{groupw}. Note that the ambient coordinates are entries of this matrix and the quantum $R$-matrix for the $\kappa$-(A)dS quantum algebra should be derived from the one associated to the Drinfel'd-Jimbo deformation~\cite{Drinfeld1987icm,Jimbo1985} of the corresponding complex simple Lie algebra. 

We would like to stress that from a physical perspective the relevant parameter appearing in the $\kappa$-(A)dS 3-space~\eqref{quantumsu2} is just $\eta/\kappa$, which is actually very small. This fact could preclude the need of considering higher order terms in the algebras~\eqref{eq:Poisson-kappaAdSst1}-\eqref{eq:Poisson-kappaAdSss1} for all physically relevant purposes. Therefore, the noncommutative algebra~\eqref{ncstime}  should suffice in order to provide  the essential information concerning the novelties introduced by the $\kappa$-(A)dS spacetime with respect to the $\kappa$-Minkowski one. In particular, the changes introduced by the cosmological constant in the representation theory of the latter~\cite{Agostini2007jmp,LMMP2018localization} are worth studying as a first step, and we recall that the irreducible representations for a complex $C^\ast$-version of the algebra~\eqref{quantumsu2} were presented in~\cite{VaksmanSoibelman1988}   (see also~\cite{ChaichianDemichev1996book}).


\section{Concluding remarks}

The construction of noncommutative (A)dS spacetimes with quantum group symmetry provides new tools aimed to explore the quantum geometry of spacetime at cosmological scales. Nevertheless, the explicit obtention of both the quantum  (A)dS  groups and their corresponding noncommutative spacetimes is, in general, a very cumbersome task, which can be illustrated by the fact that only recently the complete   $\kappa$-deformation for the (A)dS algebra in (3+1) dimensions has been achieved in~\cite{BHMN2017kappa3+1}. The aim of the present work is just the construction and analysis of the associated $\kappa$-(A)dS noncommutative spacetime by making use of a semiclassical approach based on the corresponding Poisson-Lie $\kappa$-(A)dS group and its Poisson homogeneous space.

Under the only hypothesis that the time translation generator is primitive after deformation, the uniqueness of the generalization of the $\kappa$-Poincar\'e quantum deformation to the (A)dS case has been shown, and such quantum (A)dS algebra turns out to be just the one presented in~\cite{BHMN2017kappa3+1}. Therefore, the corresponding (A)dS Poisson noncommutative spacetime can explicitly be  constructed and arises as a nonlinear deformation of the $\kappa$-Minkowski spacetime in terms of the cosmological constant parameter $\r=\sqrt{-\Lambda}$. The quantization at first-order in $\r$ of the $\kappa$-(A)dS Poisson algebra gives rise to a noncommutative spacetime which is identical to the $\kappa$-Minkowski spacetime for the time-space sector and presents a novel noncommutativity of the space sector ruled by the cosmological constant and arising from the quantum ${\rm SU}(2)$ subalgebra that exists within the $\kappa$-(A)dS deformation. Moreover, the full $\kappa$-(A)dS spacetime can be obtained in all orders in $\r$ by working in ambient space coordinates which, in turn, give rise to the $\kappa$-(A)dS spacetime as a noncommutative pseudosphere.

It is also worth mentioning that the noncommutative spacetime coming from the  $r$-matrix~\eqref{eq:r_firstfamily_twist_j3} of the twisted $\kappa$-deformation of the (A)dS group can be obtained by following the very same approach, since it again provides  a coisotropic Lie bialgebra for the Lorentz sector. As it happened with the twisted $\kappa$-Minkowski spacetime~\eqref{eq:PoissonMinkowski_twist}, the computation of the Sklyanin bracket shows that the twist does not affect the Poisson brackets between space coordinates---which are again~\eqref{eq:Poisson-kappaAdSss}---and the twisted brackets involving $x^0$ and the space coordinates $x^a$ are given by
\begin{align}
\begin{split}
\label{eq:Poisson-kappaAdStwisted}
&\{x^0,x^1\} =-\frac{1}{\kappa}\,\frac{\tanh (\r x^1)}{\r \cosh^2(\r x^2) \cosh^2(\r x^3)} - \t \,\frac{\cosh (\r x^1) \tanh (\r x^2)}{\r} ,\\
&\{x^0,x^2\} =-\frac{1}{\kappa}\,\frac{\tanh (\r x^2)}{\r \cosh^2(\r x^3)} + \t \,\frac{\sinh (\r x^1)}{\r} ,\\
&\{x^0,x^3\} =-\frac{1}{\kappa}\,\frac{\tanh (\r x^3)}{\r} \, ,
\end{split}
\end{align} 
which   again provide a nonlinear algebra deformation of the twisted $\kappa$-Minkowski whose zeroth-order in $\r$ leads to~\eqref{eq:PoissonMinkowski_twist}, and where $\t$ is the twist parameter.

Some words concerning the (2+1)-dimensional counterpart of the results here presented are in order, since it is well-known that the $\kappa$-(A)dS deformation leads to a vanishing commutation rule $[\hat x^1,\hat x^2]=0$  for the space coordinates (see~\cite{BHMN2014sigma}). This can easily be  explained by taking into account that in (2+1) dimensions the $\kappa$-(A)dS $r$-matrix reads 
\be
r_\L= \frac{1}{\kappa} \left( K_1 \wedge P_1 + K_2 \wedge P_2 \right),
\ee
and the term $J_1\wedge J_2$ which generates the space-space noncommutativity~\eqref{quantumsu2} in (3+1) dimensions cannot exist.

Indeed, the consequences of making use of the $\kappa$-(A)dS noncommutative spacetime~\eqref{ncstime} from different quantum gravity perspectives have to be explored promptly  by following similar approaches to the ones used so far for the $\kappa$-Minkowski spacetime (see, for instance,~\cite{AM2000waves}-\cite{BGH2019worldlinesplb}). Work on this line is currently in progress and will be presented elsewhere.


\section*{Acknowledgements}

This work has been partially supported by Ministerio de Ciencia, Innovaci\'on y Universidades (Spain) under grant MTM2016-79639-P (AEI/FEDER, UE), by Junta de Castilla y Le\'on (Spain) under grant BU229P18 and by the Actions MP1405 QSPACE and CA18108 QG-MM from the European Cooperation in Science and Technology (COST). I.G-S. acknowledges a predoctoral grant from Junta de Castilla y Le\'on and the European Social Fund. The authors are indebted to G. Gubitosi for useful discussions.

\end{document}